\pdfoutput=1
\documentclass{article}

\usepackage[final]{nips_2017}

\usepackage[utf8]{inputenc} 
\usepackage[T1]{fontenc}    
\usepackage{hyperref}       
\usepackage{url}            
\usepackage{booktabs}       
\usepackage{amsfonts}       
\usepackage{nicefrac}       
\usepackage{microtype}      
\usepackage{amsmath,graphicx}
\usepackage{adjustbox}      
\title{Audio Cover Song Identification using Convolutional Neural Network}

\author{
  Sungkyun Chang$^{1,4}$, Juheon Lee$^{2,4}$, Sang Keun Choe$^{3,4}$ and Kyogu Lee$^{1,4}$  \\
  Music and Audio Research Group$^1$,\\ College of Liberal Studies$^2$,\\ Dept. of Electrical and Computer Engineering,$^3$\\
  Center for Superintelligence$^4$,\\
  Seoul National University\\
  \texttt{\{rayno1, juheon2, hana9000, kglee\}@snu.ac.kr} \\
}

\begin{document}

\maketitle

\begin{abstract}
In this paper, we propose a new approach to cover song identification using a CNN (convolutional neural network). Most previous studies extract the feature vectors that characterize the cover song relation from a pair of songs and used it to compute the (dis)similarity between the two songs. Based on the observation that there is a meaningful pattern between cover songs and that this can be learned, we have reformulated the cover song identification problem in a machine learning framework. To do this, we first build the CNN using as an input a cross-similarity matrix generated from a pair of songs. We then construct the data set composed of cover song pairs and non-cover song pairs, which are used as positive and negative training samples, respectively. The trained CNN outputs the probability of being in the cover song relation given a cross-similarity matrix generated from any two pieces of music and identifies the cover song by ranking on the probability. Experimental results show that the proposed algorithm achieves performance better than or comparable to the state-of-the-art.
\end{abstract}

\section{Introduction}
\label{intro}
In popular music, a cover song or cover version is defined as a new recording produced by someone who is not an original composer or singer. Cover songs share key musical elements, such as melody contours, basic harmonic progressions, and lyrics, with the original song. However, they can differ from the original song in other aspects, such as instrumentation, tempo, rhythm, key, harmonization, and arrangement. Applications of cover song identification include content-based music recommendation, detection of music plagiarism, and music sampling, to name a few. 

Conventional methods for cover song identification generally combines a feature extraction and a distance metric. For feature extraction,  chroma feature (\citet{serra2009cross}) and its variants (\citet{muller2006towards,muller2010towards}) have been widely used for characterizing melodies and harmonic progressions. A distance metric then measures the similarity of sub-sequences in the feature space within two pieces of music. Various distance metrics, including dynamic time warping (DTW; \citet{serra2008chroma}) cost, cross-correlation (\citet{ellis20072007}), and recently, similarity matrix profile (SimPLe; \citet{silva2016simple}) and structural similarity (\citet{cai2017cross})-based methods, have been proposed for this purpose. 

So far, there have been a few attempts to exploit machine learning for cover song identification. \citet{humphrey2013data} used sparse coding with 2-dimensional Fourier magnitude coefficient derived from chroma. Recently, \citet{heocover} attempted to apply metric learning (\citet{davis2007information}) to results from SimPLe. Both of these works were based on existing rule-based cover song identification algorithms, and they mainly focused on improving the scalability of cover song discovery by proposing a novel embedding technique or metric subspace learning for the distance calculation, respectively. 

In this research, we propose a convolutional neural network-based system for audio cover song identification. We use a cross-similarity matrix generated from a pair of songs as an input feature. This idea is based on the observation that similar sub-sequences within cover songs often appear as a meaningful pattern in the cross-similarity matrix. With this assumption, we reformulate the audio cover song identification problem in the image classification framework.

\section{Basic Idea}
\label{sec:idea}
In various previous works on audio matching, the local chroma energy distributions across a shifting time window have been widely used as a representation of pitch contents, including melody contour and chord progression. Based on \citet{hu2003polyphonic}, we first convert the audio signals for each song into a 12-dimensional chroma feature with a 1 s non-overlapping window. Then, we can define a cross-similarity matrix $S$ with respect to a pair of two chroma features $\{A,B\}$ as  
\begin{equation}
S_{l,m}=\frac{\text{max}(\Delta)-\Delta_{l,m}}{\text{max}(\Delta)},\hspace*{.2cm} \text{s.t.}\hspace*{.15cm} \Delta_{\{l,m|l\in L, m\in M\}}= \delta(A_{(:,l)},B_{(:,m)}), 
\label{eq:ssm_gen}
\end{equation}
where $\delta$ denotes a distance function, and $\{L,M\}$ are the entire time indices of chroma sequence $\{A\in \mathbb{R}^{12  \times L},B\in \mathbb{R}^{12  \times M}\}$, respectively. For $\delta$, we calculate the Euclidean distance after applying the key alignment algorithm proposed in \citet{serra2008transposing}. This is also known as the optimal transposition index.

\begin{figure}[t] 
\begin{center}
\includegraphics[width=1.0\linewidth]{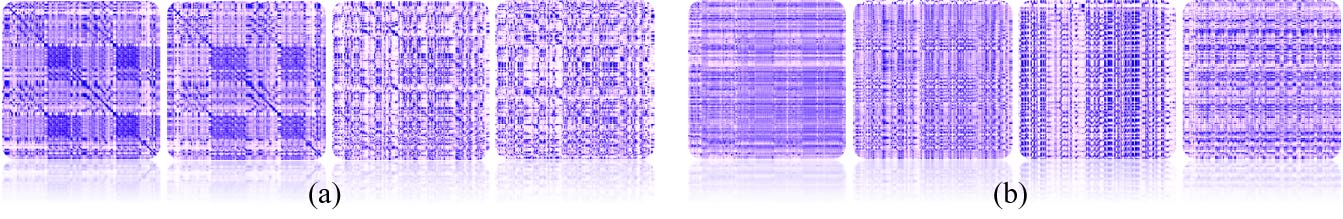}
\end{center}
\caption{Sampling of $180 \times 180$ cross-similarity matrices generated using the first 180 s of each song. (a) were generated from the cover pairs of one original song ``Toy - Passionate goodbye'' and its four different cover versions. (b) were generated using the same original song and its four different non-cover pair.} 
\label{fig:xsimmtx_ex}
\end{figure}

Fig. \ref{fig:xsimmtx_ex} displays eight examples of the $S$ generated from (a) the four cover pairs and (b) four non-cover pairs. The leftmost two images of (a) were generated from cover pairs containing almost same accompaniments, and we could observe consistent diagonal stripes with block patterns. In the third and fourth leftmost images of (a) were generated from the cover pairs produced in different tempo and instrumentations. Although the block patterns disappeared, we could observe consistent diagonal stripes in contrast with (b) from the non-cover pairs. Based on this observation, we assume that a convolutional neural network model for image classification can distinguish relevant patterns from the cross-similarity matrix. More specifically, a block of convolutional layers can sequentially perform sub-sampling and cross-correlation (or convolution) for distinguishing meaningful patterns from images in many different scales. Currently, we only compare the first 180 s of each song: We observed that most of popular music recordings had durations of three to five minutes, and  the first three minutes mostly contains main melodies. Thus, we assume that the first 180 s of each song could provide relevant information to identify a cover song. If the song lasted for less than 180 s, the duration of the song was standardized with zero-padding.

Note that Eq. \ref{eq:ssm_gen} is equivalent to the intermediate process of SimPLe proposed in \cite{silva2016simple}. Another closely related work is \citet{sakoe1978dynamic}, which exploits a cross-similarity matrix in the early process of speech alignment. In addition, similar ideas of utilizing stripe or block-like patterns in a self-similarity matrix have been proposed  in various works for audio music segmentation (\citet{paulus2010state}). All these findings motivated us to use the cross-similarity matrix with a convolutional neural network.

\section{Proposed System}
\label{sec:proposed_sys}

\begin{figure}[t] 
\begin{center}
\includegraphics[width=1.0\linewidth]{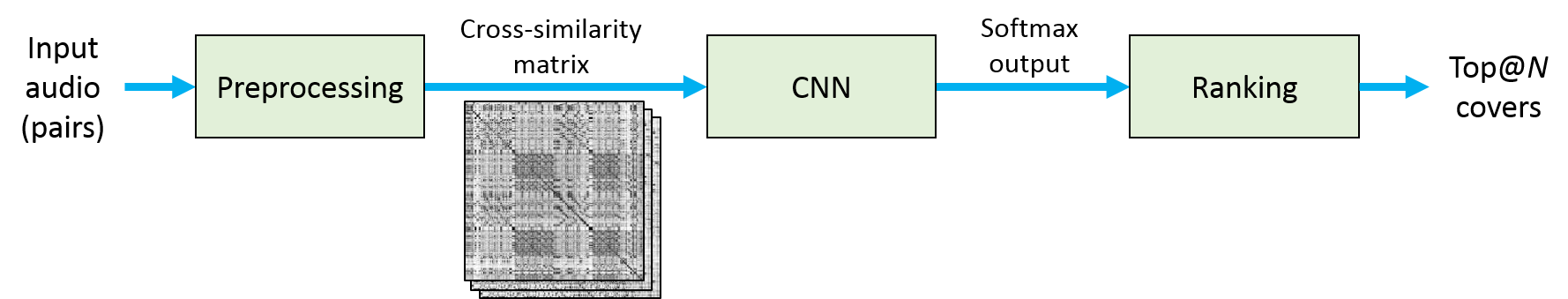}
\end{center}
\caption{Overview of the proposed system.} 
\label{fig:system_overview}
\end{figure}

The proposed system, shown in Fig. \ref{fig:system_overview}, consists of three stages. In the preprocessing stage, we convert audio signals into chroma features for each song. Then we generate cross-similarity matrices by taking a pair of chroma features as described in Section \ref{sec:idea}. 

\begin{table}[t]
\caption{Specification of convolutional neural network: Inside the brackets are unit convolutional blocks, and outside the brackets is the number of stacked blocks. {\it Conv} denotes a {\it same} convolution layer with stride = 1, and its inside parentheses is (channel$\times$width$\times$height). {\it Maxpool} denotes a max-pooling layer with stride = 1, and its inside parentheses is (pooling size). {\it BN and FC} denote batch normalization and fully-connected layer, respectively.}
\centering
\begin{adjustbox}{max width=\textwidth}
\begin{tabular}{|c|c|c|c|c|c|c|c|l|}
\hline \rule{0pt}{10pt}
Block \# & Input layer & block 1 & block 2 & block 3 & block 4 & block 5 & \multicolumn{2}{c|}{Final layers} \\ \hline \rule{0pt}{30pt}
\begin{tabular}[c]{@{}c@{}}Compo-\\ nents\end{tabular} & - & 
\begin{math}

\left \{
  \begin{tabular}{c}
  {\it Conv} (32 $\times$ 5 $\times 5$), ReLU \\
  {\it Conv} (32 $\times$ 5 $\times 5$), ReLU \\
  {\it Maxpool} (2 $\times$ 2) \\
  {\it BN}
  \end{tabular} 
\right \} \times 1

\end{math}
& \multicolumn{4}{c|}{
\begin{math}

\left \{
  \begin{tabular}{c}
  {\it Conv} (32 $\times$ 3 $\times 3$), ReLU \\
  {\it Conv} (16 $\times$ 3 $\times 3$), ReLU \\
  {\it Maxpool} (2 $\times$ 2) \\
  {\it BN}
  \end{tabular} 
\right \} \times 4

\end{math}

} &
\begin{tabular}{c}
  {\it DropOut}$^p$(0.5)\\
  {\it FC(256)}, ReLU\\
  {\it DropOut}$^q$(0.25)
  
\end{tabular} 

& \begin{tabular}[c]{@{}l@{}}{\it FC}(2)\\ softmax\end{tabular} \\ \hline
\begin{tabular}[c]{@{}c@{}}Output \\ shape\end{tabular} & (1, 180, 180) & (32,90,90) & (16,45,45) & (16,22,22) & (16,11,11) & (16,5,5) & (,,256) & \multicolumn{1}{c|}{(,,2)} \\ \hline
\end{tabular}
\end{adjustbox}
\label{tab:cnn_spec}
\end{table}

The next stage is based on the convolutional neural network (hereafter, CNN), as specified in Table \ref{tab:cnn_spec}. Our CNN is built as a narrower and deeper network ($0.58 \times$ $10^6$ parameters with 10 convolution layers) than conventional CNNs for ImageNet, such as AlexNet(\citet{krizhevsky2012imagenet}) which has $60\times 10^6$ parameters with five convolution layers. With respect to the size of the input cross-similarity matrix, we currently fix it as $180\times 180$ (cut or zero-padded) that corresponds to comparing the first 3 min of music. With respect to the filter size of the first convolution layer, the receptive field of the first layer corresponds to 5 s of audio (2--4 measures in a music score). In practice, using the first convolution filter size of $5\times 5$ resulted in approximately 4 \% better performance than using $3\times 3$ or 7$\times$7.  With respect to blocks 2--4, the basic idea in Section \ref{sec:idea} was to run a chain of processing pattern consisting of sub-sampling and cross-correlation (or convolution) with these blocks. For this, blocks 2-4 of the CNN are built using a template convolutional block that outputs a one-half down-sampled size. In every convolutional block,  we apply batch normalization (\citet{ioffe2015batch}). 

The last stage of our system performs ranking on the softmax output of the trained CNN. We first take the cover-likelihood vector over all cover candidates. Then we apply descending-sort on this vector for ranking the most likely top@$N$ covers.

\section{Experimental Results}
\label{sec:Results}

\subsection{Data set}
\label{subsec:dataset}
We use an evaluation data set provided by \citet{heocover}. The data set resembles that used for the MIREX\footnote{\url{http://www.music-ir.org/mirex/wiki/}} cover song identification task. It consists of 330 cover songs that make the query set, and 670 dummy songs that are not covered. Of the 330 query songs, there are 30 different kinds of cover songs. Each has 11 different cover versions (each query song must have 10 ground-truth covers). Thus, it can yield test examples of 3,300 cover pairs and 496,200 non-cover pairs.
 
The training set consists of 2,113 cover pairs and 2,113 non-cover pairs. The held-out validation set consists of 322 cover pairs and 322 non-cover pairs. These data sets are disjoint. The audio files contain popular Korean music released from 1980 to 2016. They were produced in stereo with a sampling rate of 44,100 Hz.

\subsection{Training}
\label{subsec:training}
In advance of the training, we applied zero-mean unit standardization on the input cross-similarity matrices for feature scaling. We trained the CNN with a total of 4,226 cross-similarity matrices (class-balanced for cover and non-cover) . The CNN was implemented based on the Keras framework, and run on a single GPU cloud server. Using the Adam optimizer (\citet{kingma2014adam}), the training stopped when the cross entropy loss $\epsilon$ reached convergence for $\epsilon < 10^{-4}$. Using a nested grid-search, we tried to optimize the two drop-out hyperparameters, denoted as drop-out$^p$ and drop-out$^q$ in Table \ref{tab:cnn_spec}. We achieved the final validation accuracy 83.4\% for drop-out$^p$ (0.5) and drop-out$^q$ (0.5), by not looking at test set accuracy.

\subsection{Results}
\label{subsec:results}
We evaluated the proposed system by following metrics proposed in the MIREX for audio cover song identification task: 
\begin{itemize}
\item MNIT10: mean number of covers identified in top 10. 
\item MAP: mean average precision.
\item MR1: mean rank of the first correctly identified cover.
\end{itemize}
Here, MNIT10 was calculated as \{total number of correctly identified covers in top 10\} divided by \{total number of ground-truth covers (= 3,300)\}.
\begin{table}[h]
\centering
\caption{Performance of audio cover song identification}
\label{tab:performance}
\begin{tabular}{@{}llll@{}}
\toprule
Model                  & MNIT10 & MAP  & MR1  \\ \midrule
SimPLe \citet{silva2016simple}  & 6.8    & 0.66 & 5.6 \\
SimPLe + Metric Learning (\citet{heocover}) & 7.9    & 0.81 & 15.1 \\
CNN (proposed)         & 8.04   & 0.84 & 2.50
\end{tabular}
\end{table}

In Table \ref{tab:performance}, we compared our system with two baseline algorithms: \citet{silva2016simple}, a rule-based algorithm, and \citet{heocover}, a metric learning-based algorithm. The largest MNIT10 was achieved by the proposed CNN. This implies that the search result of the proposed system contained 8.04 correct covers out of 10, in average. With respect to MNIT10 and MAP (where larger is better), the present CNN showed competitive precision over the two compared algorithms. With respect to MR1 (where smaller is better), the proposed CNN achieved 80.10\% improved performance over SimPLe, the second-best algorithm. The smaller MR1 implies that the ground-truth covers would more consistently appear in top search results. The effect of comparing various input lengths of each song has not been examined yet. However, the proposed system comparing only the first 180 s achieved the better performance over than all other systems comparing the entire lengths of input songs.

\section{Conclusions and Future Work}
We proposed a convolutional neural network-based approach to audio cover song identification. Our assumption was that the cross-similarity matrix from a pair of two songs could appear as a meaningful pattern. Based on this, we trained the CNN using cross-similarity matrices in the same manner that a binary classifier for images is trained. By ranking the softmax output from the trained CNN, the proposed system was able to predict a fixed number of the most likely cover song pairs. The performance of the proposed system was compared with a rule-based approach and another machine learning-based approach. Although the current study showed promising results, there is much room for improvement, particularly by finding more a suitable CNN design, hyper-parameter tuning, and increasing the size of the training data set with flexible input feature length. Furthermore, we did not apply any of the embedding techniques that are necessary for a large-scale search of cover songs. Thus, exploration of these is left for future work.

\subsubsection*{Acknowledgments}
This work was supported by Kakao and Kakao Brain
corporations.
\bibliographystyle{unsrtnat}
\bibliography{bib.bib}

\end{document}